\documentclass[10pt,conference]{IEEEtran}
\IEEEoverridecommandlockouts
\usepackage{cite}
\usepackage{amsmath,amssymb,amsfonts}
\usepackage{algorithmic}
\usepackage{graphicx}
\usepackage{textcomp}
\usepackage{xcolor}
\usepackage{hyperref}
\usepackage{tabularx}
\usepackage{multirow}
\usepackage{subcaption}
\usepackage{pifont}
\usepackage{balance} 
\usepackage{enumitem}

\usepackage{array}
\newcolumntype{L}[1]{>{\raggedright\let\newline\\\arraybackslash\hspace{0pt}}m{#1}}
\newcolumntype{C}[1]{>{\centering\let\newline\\\arraybackslash\hspace{0pt}}m{#1}}
\newcolumntype{R}[1]{>{\raggedleft\let\newline\\\arraybackslash\hspace{0pt}}m{#1}}

\def\BibTeX{{\rm B\kern-.05em{\sc i\kern-.025em b}\kern-.08em
    T\kern-.1667em\lower.7ex\hbox{E}\kern-.125emX}}
\begin{document}

\title{Recommender Systems for Sustainability:\\ Overview and Research Issues}

\author{
\IEEEauthorblockN{1\textsuperscript{st} Alexander Felfernig}
\IEEEauthorblockA{\textit{Graz University of Technology, Austria}\\
alexander.felfernig@tugraz.at}
\and
\IEEEauthorblockN{2\textsuperscript{nd} Manfred Wundara}
\IEEEauthorblockA{\textit{Magistrat Villach, Villach, Austria}\\
manfred.wundara@student.tugraz.at}
\and
\IEEEauthorblockN{3\textsuperscript{rd} Thi Ngoc Trang Tran}
\IEEEauthorblockA{\textit{Graz University of Technology, Austria}\\
ttrang@tugraz.at}
\and
\IEEEauthorblockN{4\textsuperscript{th} Seda Polat Erdeniz}
\IEEEauthorblockA{\textit{Graz University of Technology, Austria}\\
sedapolat@gmail.com}
\and
\IEEEauthorblockN{5\textsuperscript{th} Sebastian Lubos}
\IEEEauthorblockA{\textit{Graz University of Technology, Austria}\\
sebastian.lubos@tugraz.at}
\and
\IEEEauthorblockN{6\textsuperscript{th} Merfat El Mansi}
\IEEEauthorblockA{\textit{Graz University of Technology, Austria}\\
merfat.el-mansi@student.tugraz.at}
\and
~~~~~~~~~~~~~~~~~~~~~\IEEEauthorblockN{7\textsuperscript{th} Damian Garber}
\IEEEauthorblockA{~~~~~~~~~~~~~~~~~~~~~\textit{Graz University of Technology, Austria}\\~~~~~~~~~~~~~~~~~~~~~
damian.garber@tugraz.at}
\and
\IEEEauthorblockN{8\textsuperscript{th} Viet-Man Le}
\IEEEauthorblockA{\textit{Graz University of Technology, Austria}\\
v.m.le@tugraz.at}
}

\maketitle
\begin{abstract}

Sustainability development goals (SDGs) are regarded as a universal call to action with the overall objectives of planet protection, ending of poverty, and ensuring peace and prosperity for all people. In order to achieve these objectives, different AI technologies play a major role. Specifically, recommender systems can provide support for organizations and individuals to achieve the defined goals. Recommender systems integrate  AI technologies such as machine learning, explainable AI (XAI), case-based reasoning, and constraint solving in order to find and explain user-relevant alternatives from a potentially large set of options. In this article, we summarize the state of the art in applying recommender systems to support the achievement of sustainability development goals. In this context, we discuss  open issues for future research.
\end{abstract}

\section{Introduction}\label{section:introduction}

The overall objective of the 17 sustainability development goals (SDGs -- see Table \ref{tab:SDGOverview}) (e.g., \emph{no poverty} and \emph{quality education}) is to provide a universal call to \emph{end poverty}, \emph{planet protection}, and to \emph{ensure that people enjoy peace and prosperity} also with the goal to establish a balance of social, economic, and environmental sustainability.\footnote{https://www.undp.org/sustainable-development-goals} Existing research \cite{vanWynsberghe2021} has already shown that Artificial Intelligence (AI) methods and techniques can have positive as well as negative impacts ranging from efficient energy production and distribution to negative aspects such as increasing power consumption scenarios due to different types of large-scale machine learning efforts \cite{Vinuesa2020}. In this article, we analyze potentials of recommender systems as a key technology to support the mentioned SDGs. 

Recommender systems can be regarded as decision support systems combining AI technologies such as machine learning, explanations, and intelligent user interfaces with the overall goal to improve a user`s decision quality \cite{Bui2000,FalknerFelfernigHaag2011}.  There are different types of recommender systems with differing applicability depending on the underlying recommendation scenario. (1) \emph{Collaborative filtering}  (CF) \cite{Ekstrand2011CFBook} follows the idea of word-of-mouth promotion where opinions of family members and friends (the so-called "nearest neighbors") are regarded as relevant recommendations for a person. (2) \emph{Content-based Filtering} (CBF) \cite{Pazzani2007} is based on the idea that if a person had specific preferences in the (near) past, these preferences would more or less remain stable and can be used for future item recommendations. (3) \emph{Knowledge-based recommender systems} (KBR) \cite{Burke2000}  are based on the idea of determining recommendations on the basis of a more in-depth semantic knowledge expressed, for example, in terms of constraints \cite{FelfernigBurke2008} or with attribute-level similarity metrics \cite{ChenPuCritiquingRecSys2012}. (4) \emph{Hybrid recommender systems} (HYB) \cite{Burke2002HybridRecSys} focus on exploiting synergy effects by trying to combine the advantages of different recommendation approaches, for example, combining CF and CBF helps to tackle the challenges of ramp-up problems (when, e.g., CF rating data are not available for a specific user). (5) \emph{Group recommender systems} (GRP) \cite{FelfernigGroupRecommenderSystems2018} focus on the determination of recommendations for groups, i.e., not individual users. Such approaches have to identify recommendations that help to achieve -- in one way or another -- a consensus among group members.\footnote{Further details on technical backgrounds of these recommendation approaches will be provided in examples introduced in Section \ref{section:recsyssustainability}.} 

In this article, we focus on indicating in which ways recommender systems can be applied to better achieve the mentioned SDGs. With this, the major contributions of our article are the following: (1) we provide an overview of the current state-of-the-art in applying recommender systems for achieving the 17 SDGs. (2) on the basis of this overview, we discuss different open issues for future research. (3) For the given SDGs, we provide concrete working examples of how to apply recommender systems. The contributions of this article enhance existing topic-related overviews \cite{Bui2000,vanWynsberghe2021,Vinuesa2020} in terms of (1) a  focus on recommender systems technologies for sustainability, (2) the provision of concrete examples of how recommender systems can be applied to achieve individual SDGs, and (3) a discussion of recommender systems specific open research issues. 

Basic insights from this overview can be summarized as follows. (1) recommender systems can already be regarded as an important technology to support the achievement of sustainability development goals. For each of the existing SDGs, corresponding recommender approaches could be identified. (2) although an application majority of CF recommenders could be observed, all of mentioned recommendation approaches (CF, CBF, KBR, HYB, and GRP) have sustainability-related applications. (3) for the discussed recommender applications, two different levels of recommender "users" exist: first, a \emph{macro-level} with more abstract organizations (e.g., countries) and second, a \emph{micro-level} with concrete entities (e.g., citizens). 

The remainder of this article is organized as follows. In Section \ref{section:methodology}, we present our methodological approach to analyze and summarize the existing state of the art in applying recommender systems to achieve sustainability development goals (SDGs). Section \ref{section:recsyssustainability} provides an overview of the 17 SDGs  and a detailed overview of the current state of the art in applying recommender systems for achieving these goals. From this discussion of the existing best-practices, we summarize related open issues for future research (see Section \ref{section:researchissues}). Finally, this article is concluded within Section \ref{section:conclusions}.


\section{Methodology}\label{section:methodology}
In this article, we focus on a comprehensive overview of the existing state of the art in \emph{recommender systems for sustainability}. Based on the gained insights, we discuss application potentials and related open issues for future research. Our analysis of the state of the art is based on a literature review with the related phases of selecting potentially relevant papers, reviewing those papers, and a discussion of the identified papers with regard to relevance for this overview article. Paper identification is based on querying existing leading research platforms with topic-related keywords. Thereafter, the identified papers have been classified with regard to their inclusion in this overview article. In this context, queries have been performed on (1) the research platforms Google Scholar\footnote{https://scholar.google.com/}, ResearchGate\footnote{https://www.researchgate.net/}, ScienceDirect\footnote{https://www.sciencedirect.com/}, SpringerLink\footnote{https://link.springer.com/}, and Elsevier\footnote{https://www.elsevier.com/}, IEEE, \footnote{https://www.ieee.org/}, and ACM \footnote{https://www.acm.org/} and (2) recommender systems related conferences and journals including ACM Recommender Systems (ACM RecSys), ACM User Modeling and User-Adapted Interaction (ACM UMAP), ACM Intelligent User Interfaces (ACM IUI), and ACM SIGCAS/SIGCHI  Computing and Sustainable Societies (COMPASS). In this context, we used the initial search queries (and different combinations thereof) of "recommender systems" + "sustainability" + "sustainability goals" + "artificial intelligence" + "decision support". Using the snowballing technique \cite{Wohlin2014}, we analyzed further topic-relevant references starting with the original set of identified papers. Overall, we have identified  122 relevant papers which served as a basis for writing this overview.

\section{Recommender Systems for Sustainability}\label{section:recsyssustainability}

In contrast to existing approaches to evaluate the impact of recommender systems which are primarily focused on different  e-commerce scenarios \cite{JannachJugovac2019}, we focus on the impact of recommender systems in terms of achieving  sustainability development goals -- Table \ref{tab:SDGOverview} provides a short overview of the 17 United Nations (UN) sustainability development goals. In the following discussions, we differentiate between  (1) a \emph{macro-level} representing recommendations determined for abstract organizations (e.g., countries, company types, and types of study programmes) and (2) a \emph{micro-level} representing recommendations determined for concrete entities (e.g., citizens, companies, and tourists). We exemplify the application of recommender systems with a focus on basic recommendations approaches, i.e., the goal in this article is to discuss application scenarios but not primarily detailed algorithmic approaches.

\begin{table*}[ht]
\centering \caption{An overview of the United Nations Sustainable Development Goals  (SDGs $1-17$).}
\begin{tabular}{|C{0.55cm}||C{4.0cm}|C{11.5cm}|} 
\hline
  ID & SDG & Description                                   \tabularnewline  \hline
 1 & No Poverty & End poverty in all its forms everywhere \tabularnewline  \hline
 2 & Zero Hunger & End hunger, achieve food security and improved nutrition and promote sustainable agriculture \tabularnewline  \hline
 3 & Good Health and Wellbeing & Ensure healthy lives and promote well-being for all at all ages \tabularnewline  \hline
 4 & Quality Education & Ensure inclusive and equitable quality education and promote lifelong learning opportunities for all \tabularnewline  \hline
 5 & Gender Equality & Achieve gender equality and empower all women and girls \tabularnewline  \hline
 6 & Clean Water and Sanitation & Ensure availability and sustainable management of water and sanitation for all \tabularnewline  \hline
 7 & Affordable and Clean Energy & Ensure access to affordable, reliable, sustainable and modern energy for all \tabularnewline  \hline
 8 & Decent Work and Economic Growth & Promote sustained, inclusive and sustainable economic growth, full and productive employment and decent work for all \tabularnewline  \hline
 9 & Industry, Innovation, and Infrastructure & Build resilient infrastructure, promote inclusive and sustainable industrialization and foster innovation \tabularnewline  \hline
 10 & Reduced Inequalities & Reduce inequality within and among countries \tabularnewline  \hline
 11 & Sustainable Cities and Communities & Make cities and human settlements inclusive, safe, resilient and sustainable \tabularnewline  \hline
 12 & Responsible Consumption and Production & Ensure sustainable consumption and production patterns \tabularnewline  \hline
 13 & Climate Action & Take urgent action to combat climate change and its impacts \tabularnewline  \hline
 14 & Life below Water & Conserve and sustainably use the oceans, seas and marine resources for sustainable development \tabularnewline  \hline
 15 & Life on Land & Protect, restore and promote sustainable use of terrestrial ecosystems, sustainably manage forests, combat desertification, and halt and reverse land degradation and halt biodiversity loss \tabularnewline  \hline
 16 & Peace, Justice, and Strong Institutions & Promote peaceful and inclusive societies for sustainable development, provide access to justice for all and build effective, accountable and inclusive institutions at all levels \tabularnewline  \hline
 17 & Partnerships for the Goals & Strengthen the means of implementation and revitalize the Global Partnership for Sustainable Development \tabularnewline  \hline
\end{tabular} 
\label{tab:SDGOverview} 
\end{table*}

\subsection{No Poverty}

The related major goal is to \emph{end poverty everywhere}. Poverty has a multitude of definitions  and can be characterized in a monetary dimension in terms of not having enough money to maintain his/her livelihood -- a related overview of AI methods to estimate the degree of poverty in a region/country can be found in \cite{Usmanovaetal2022}. Examples of data sources used in such contexts are, for example, household data (e.g., demographics, education, and food consumption), food price data, and e-commerce data \cite{Usmanovaetal2022}. Poverty prediction has to be accompanied with approaches that help to counteract poverty. For example,  \cite{Che2020IntelligentED} show how recommendation techniques can be applied to identify export diversification strategies in such a way that a country has a latent competitive advantage (when following this strategy). 

An important measure in this context is the so-called  \emph{Revealed Comparative Advantage
(RCA)} score (for a country $\theta$ and product $\pi$) (see Formula \ref{eq:RCAscore}) \cite{BelassaNoland1989} which is used to determine the importance of individual items (products) in the export basket of a country. In this context, $E_{\theta\pi}$ is the export value of item (product) $\pi$ for country $\theta$.

\begin{equation}\label{eq:RCAscore}
 RCAScore_{\theta\pi} = \frac{E_{\theta\pi}/ \Sigma_{\pi}E_{\theta\pi}}{\Sigma_{\theta}E_{\theta\pi} / \Sigma_\theta\Sigma_\pi E_{\theta\pi}}  
\end{equation}

In the line of  \cite{Che2020IntelligentED}, recommendation services can be provided on the basis of the $RCAScore$ of individual items. When applying collaborative filtering (CF), an item $\times$ $RCAScore$ matrix summarizes the scores of items already exported by individual countries. CF can now be applied to predict the relevance ($RCAScore$) of new items not exported by individual countries up to now. In the example shown in Table \ref{tab:RCAScoreExample}, basic $RCAScore$ information is already available for products such as \emph{computer}, \emph{tourism}, and \emph{wine}.

\begin{table}[ht]
\centering \caption{Example: applying collaborative filtering for recommending advantageous items (products).}
\begin{tabular}{|c|c|c|c|c|c|} 
\hline
  item (product) & country$_1$ & country$_2$ & ... & country$_n$                                   \tabularnewline  \hline
computer & 1.5 & 2.2 & ... & 1.5 \tabularnewline  \hline
tourism & 1.1 & 2.8 & ... & 1.1 \tabularnewline  \hline
wine & 1.3 & 0.5 & ... & 1.2 \tabularnewline  \hline
... & ... & ... & ... & ... \tabularnewline  \hline
automotive & 3.1 & 2.2 & ... & 4.1 \tabularnewline  \hline \hline
solar equipment & ? & ? & ... & 5.1 \tabularnewline  \hline
\end{tabular} 
\label{tab:RCAScoreExample} 
\end{table}

Some countries do not export some of the products and we would like to know for which additional products (items) it would be good for a country to extend its assortment. In Table \ref{tab:RCAScoreExample}, "?" indicates that a recommendation is needed, for example, for country$_1$,  it would be good to focus on producing and exporting solar equipment. Based on the idea of CF, the nearest neighbor of country$_1$ is country$_n$ (the nearest neighbor is regarded as a country with a similar $RCAScore$ distribution) with a high relevance of exporting solar equipment. In this simplified scenario, engaging in exporting solar equipment can be regarded also as a good idea for country$_1$. For a detailed discussion of applying different CF algorithms in such application contexts, we refer to  \cite{Che2020IntelligentED}. Furthermore, \cite{Liao2018} discuss approaches to product diversification based on the concepts of social network analysis where relationships between countries and their products are analyzed for recommendation purposes.

On the level of individuals, poverty can be triggered by various factors such as wrong investment decisions (e.g., purchasing a too expensive car and dealing with the consequences), wrong choice of personal education and employment (e.g., to stop visiting school with the consequences of problems in finding a job), and issues in handling the personal financial situation (e.g., women focusing on childcare and without a corresponding financial provision). In the following, we provide a simple example of applying a knowledge-based recommendation \cite{Felfernig2006} approach as a basic support in investment decisions \cite{FanoKurthChoicePoint2003}. Table \ref{tab:portfolioattributes} provides an overview of different portfolio elements that could be selected by the user of a recommender system.

\begin{table*}[ht]
\centering \caption{Example: simplified portfolio elements (with costs per month).}
\begin{tabular}{|c|c|c|c|c|c|c|c|c|c|c|c|c|c|c|c|c|c|c|} 
\hline
  attributes       & \multicolumn{3}{|c|}{car}                & \multicolumn{2}{|c|}{house}             & \multicolumn{2}{|c|}{workers}   & \multicolumn{2}{|c|}{holidays}          & \multicolumn{2}{|c|}{food}       \tabularnewline  \hline
  domains    & BMW&Renault&none  & large&medium & 1&2       & yes&no     & flexible&restricted \tabularnewline  \hline
  costs/income      & 500&350&0        &2.5k&1.5k  & 2k&3k     & 150&0    & 600&200      \tabularnewline  \hline

\end{tabular} 
\label{tab:portfolioattributes} 
\end{table*}

A major criterion in portfolio recommendation is that the overall \emph{consumed resources} in terms of costs \emph{c}  (\emph{car}, \emph{house}, \emph{holidays}, and \emph{food} representing, e.g., family dinner etc.) must not exceed the provided resources (income provided by \emph{workers} per year). This resource limitation can be expressed as shown in Formula \ref{eq:exampleresourceconstraint} where the property \emph{w.income} (\emph{w=workers}) represents the monthly income of the family. 

\begin{equation} \label{eq:exampleresourceconstraint}
    12 \times (car.c+house.c+holidays.c+food.c) \leq 12 \times w.income
\end{equation}

On the basis of such a scenario, the user of a recommender system can choose different options, for example, an expensive car and an expensive house, and immediately understand the consequences of such decisions. For example, with the current yearly income, it is impossible to have both, an expensive car and a large house. Furthermore, there also exists a scenario (portfolio) where one worker would in principle be enough to cover all of the estimated costs. Table \ref{tab:portfolioexamples} shows the extreme cases of a portfolio with \emph{maximum costs} p.a. (45k) and the other extreme of \emph{minimum costs} p.a. (20.4k).

\begin{table}[ht]
\centering \caption{Example portfolios and associated costs p.a.}
\begin{tabular}{|c|c|c|c|c|c|c|c|c|c|c|c|c|c|c|c|c|c|c|} 
\hline
  portfolio  &   car    & house   &  holidays    & food       & total costs p.a.      \tabularnewline  \hline
  max        &   BMW    & large   &  yes         & flexible   & 45k                   \tabularnewline  \hline
  min        &   none  & medium  &  no          & restricted & 20.4k                 \tabularnewline  \hline
\end{tabular} 
\label{tab:portfolioexamples} 
\end{table}

The presented example is a simplified variant of a knowledge-based recommender system focusing on showing to the user the impacts of specific investment decisions. In situations where the defined user preferences do not allow the recommendation of a portfolio, corresponding diagnosis techniques can help to indicate minimal changes in the users preferences in such a way that a solution can be identified.\footnote{For further related details, we refer to  \cite{Felfernig2006}.}

\subsection{Zero Hunger}
The related goal is to \emph{end hunger and to achieve improved nutrition and food security while at the same time promoting sustainable agriculture}. In contrast to the application of recommender systems in the context of healthy living \cite{Tranetal2018HealthyFood}, a major focus of sustainability in the context of achieving \emph{zero hunger} is to foster more conscious food consumption and to support food production processes with a clear sustainability focus \cite{BounietalRecSysSmartAgriculture,Giletal2021,Martinietal2022GuideFoodSecurity}. 
A related crop diversification (recommendation), i.e., choosing and diversifying crops, can help governments to grow more crops in ones own country and with this to reduce dependencies to other countries \cite{Giletal2021}. This also includes mechanisms to effectively detect crop diseases \cite{Omara2023}.

The appropriate determination of crop factors such as maturity date, soil suitability, and pesticide requirements becomes increasingly important. Not least, to be able to choose the optimal crop in the long run as well as to optimize production and to minimize additional efforts in terms of pesticides and soil fertilization. A simplified example of a potential application of recommender systems in crop selection is shown in  Table \ref{tab:examplecroprecommendation}. In this example, the question is if $crop_2$ (the \emph{current} entry) could be relevant for region $D$ (no corresponding experience data available). Since average temperature and soil moisture are quite similar to region $C$ (the nearest neighbor -- \emph{id}=5), the expected crop$_2$ output for this region is about $83$\% with a recommended pesticide usage $p_3$. In real-world settings, further parameters are needed for determining high-quality recommendations \cite{Giletal2021}.

\begin{table*}[ht]
\centering \caption{Example of collaborative filtering based crop recommendation.}
\begin{tabular}{|c|c|c|c|c|c|c|c|c|c|c|c|c|c|c|c|c|c|c|} 
\hline
  id & name &   region    & pesticides   &  avg. temperature (cel.)    & soil moisture (\%) & output (\%)        \tabularnewline  \hline
  1  & crop$_1$  &   A         & $p_1$        &  20                         & 30                 & 70            \tabularnewline  \hline
  2  & crop$_1$  &   B         & $p_2$        &  22                         & 25                 & 75            \tabularnewline  \hline
  3  & crop$_2$  &   E         & $p_2$        &  22                         & 25                 & 75            \tabularnewline  \hline
  4  & crop$_1$  &   A         & $p_2$        &  20                         & 27                 & 76            \tabularnewline  \hline
  5  & crop$_2$  &   C         & $p_3$        &  20                         & 27                 & 83            \tabularnewline  \hline \hline
  current  & crop$_2$  &   D         & ?            &  20                         & 27                 & ?             \tabularnewline  \hline
\end{tabular} 
\label{tab:examplecroprecommendation} 
\end{table*}

 \emph{Food rescue organizations} focus on collecting and delivering food donations to those in need \cite{ShietalCrowdsourcingfoodrescue2021}. In many cases, collected food is in temporary storage at the rescue organization where it is offered to persons in need. Collecting the food from various local food providers is a logistic problem in the sense that volunteers need to be identified who are willing to take over a specific pick-up and food delivery task. \cite{ShietalCrowdsourcingfoodrescue2021} present a recommender system that helps to identify candidate persons with a high probability of willing to perform a new collection and delivery task. 

A simplified example of supporting such scenarios on the basis of content-based filtering  is depicted in Table \ref{tab:examplevolunteerrecommendation}. In this setting, a new collection task is defined for region $A$ and includes beverages and meat. Important to know is that many food rescue organizations allow their volunteers to claim a low share of each cartload for their own. Based on this assumption, a content-based recommender system can identify those potential drivers (volunteers) who might be interested in performing the collection task. In our example, $user_3$ can be regarded as having preferences which are most similar to those of $task_{new}$ -- consequently, $user_3$ can be regarded as the first candidate to be contacted.

\begin{table}[ht]
\centering \caption{Example of volunteer (user) recommendation with content-based filtering. Each table row represents a (simplified) user profile, for example, the entry $drinks=yes$ of $user_1$ indicates that $user_1$ prefers collection tasks with beverages included.}
\begin{tabular}{|c|c|c|c|c|c|c|c|c|c|c|c|c|c|c|c|c|c|c|} 
\hline
  user &       region    & beverages   &  meat    & bread & vegetables        \tabularnewline  \hline
  $user_1$       &   A         & yes         &  no      & yes   & no                \tabularnewline  \hline
  $user_2$       &   B         & no          &  no      & no    & yes               \tabularnewline  \hline
  $user_3$       &   A         & yes         &  yes     & yes   & no                \tabularnewline  \hline \hline
  $task_{new}$     &   A         & yes         &  yes     & no    & no                \tabularnewline  \hline
\end{tabular} 
\label{tab:examplevolunteerrecommendation} 
\end{table}

For sure, in real-world settings, further related parameters can play an important role in recommending volunteers. Examples of such parameters are \emph{availability} (a user might be available only during specific time periods), \emph{fairness} (all volunteers should have near-equal chances to be contacted), and \emph{reliability} (e.g., the driver always in-time). A detailed discussion of the application of recommender systems in a food rescue scenario is given in \cite{ShietalCrowdsourcingfoodrescue2021}.

\subsection{Good Health and Wellbeing}

The related goal is \emph{to ensure healthy lives and promote wellbeing}. The success of public health campaigns heavily depends on the appropriateness of health messages delivered to users \cite{Cappellaetal2015}. In such scenarios, recommender systems can help to personalize message delivery given some knowledge about features and topics of interest for a user. A simple approach can be a topic-wise recommendation where new messages/campaigns are forwarded to citizens in a personalized fashion. A related simplified example is depicted in  Table \ref{tab:examplepersonalizedmessaging}: user interests are stored in a corresponding user profile, for example, $user_3$ has a high interest in healthy eating and healthy cooking.  A new health campaign should be issued and the task is to identify those users with some basic potential interest in the related topics. The most relevant topics of $message_{new}$ are \emph{healthy eating} and \emph{healthy cooking} -- in this scenario $user_3$ and to some extent $user_2$ have related interests, i.e., these users should be contacted in the context of the new campaign. As such, this is a simple example of applying content-based filtering in the context of delivering public health campaigns \cite{Cappellaetal2015}. To assure that users get also in touch with new topics, diversity-enhanced and collaborative recommendation can be applied to increase serendipity effects \cite{ZiaraniRavanmehrSerendipity2021}. 

\begin{table*}[ht]
\centering \caption{Example personalized message delivery in public health campaigns.}
\begin{tabular}{|c|c|c|c|c|c|c|c|c|c|c|c|c|c|c|c|c|c|c|} 
\hline
  user            &  healthy eating    & athletic sports   &  endurance sports   & healthy cooking & sports events    \tabularnewline  \hline
  $user_1$        &   0.5              & 0.8               &  0.0                & 0.2             &  0.0           \tabularnewline  \hline
  $user_2$        &   0.5              & 0.1               &  0.6                & 0.5             &  0.8           \tabularnewline  \hline
  $user_3$        &   0.9              & 0.2               &  0.5                & 0.9             &  0.2           \tabularnewline  \hline \hline
  $message_{new}$ &   0.9              & 0.1               &  0.5                & 0.9             &  0.0           \tabularnewline  \hline 
\end{tabular} 
\label{tab:examplepersonalizedmessaging} 
\end{table*}

Another related example on the macro-level is the support of machine learning and recommender systems in the context of vaccine allocation and distribution where appropriate planning and fairness aspects play a major role \cite{Blasiolietal2023}. In this scenario, aspects such as population size, percentage of individuals who have already received a previous dose, and storage capacity for the vaccines are important factors to be taken into account. An overview of the application of recommender systems in the healthcare domain is provided, for example, in \cite{Tranetal2018}. Important to mention, related applications are quite diverse and not all of those can be discussed in this article. Examples of recommender systems in the healthcare domain range from healthy food recommendation \cite{Wang2021}, personal wellbeing \cite{Arevaloetal2022AirPollutionOutdoorActivities2022}, air pollution  aware outdoor activity recommendation \cite{AlcarazHerrera2022}, context-aware sleep health recommenders \cite{Liang2022}, context-aware recommenders for diabetes patients \cite{AbuIssaetal2023}, activity recommenders for elderly \cite{Herpichetal2017}, to the recommendation of  healthcare professionals \cite{Singh2023}.

A simplified example of an approach to recommend food items in a healthiness-aware fashion (and -- at the same time -- to take into account food preferences of the current user) is apply collaborative filtering for selecting food items and then to filter relevant items using a knowledge-based approach. Table \ref{tab:examplehealthyfoodrecommendation} depicts a collection of recipes (for simplicity, we assume main dishes) and corresponding user preferences. The \emph{current} user has already consumed \emph{schnitzel} and \emph{lasagne} in the past. A recommender could recommend these or similar items also in the future  (e.g., \emph{veal}). However, since both selections have rather low nutritional values \cite{Fialon2021}, an alternative is to recommend \emph{salad} and \emph{spinach} which has also been consumed by the nearest neighbor \emph{$user_1$}. 

\begin{table*}[ht]
\centering \caption{Example food item consumption with corresponding front-of-pack labels ($a$ ..$e$) where $a$ indicates high and $b$ low nutritional values \cite{Fialon2021}.}
\begin{tabular}{|c|c|c|c|c|c|c|c|c|c|c|c|c|c|c|c|c|c|c|} 
\hline
  user           & schnitzel$_e$ & beans$_a$ & soja$_b$  & veal$_d$ & lasagne$_c$ & trout$_b$  & spaghetti$_b$ & spinach$_a$ & salad$_a$     \tabularnewline  \hline
  $user_1$       & x         &       &       & x    & x       &        &           &   x      &   x       \tabularnewline  \hline
  $user_2$       &           &  x    &  x    &      &         &   x    &           &   x     &   x       \tabularnewline  \hline
  $user_3$       &           &       &  x    &      &         &        &           &         &    x      \tabularnewline  \hline \hline
$current$        & x         &  ?    &  ?    &  ?   &  x       &  ?      &     ?     &    ?    &    ?      \tabularnewline  \hline
\end{tabular} 
\label{tab:examplehealthyfoodrecommendation} 
\end{table*}

The idea of such a recommender could be to create diversity in terms of identifying items (or recipes) the current user did not consume up to now and -- at the same time -- to take into account nutritional values, i.e., to prefer items with high nutritional values (e.g., \emph{salad} or \emph{spinach}). Just recommending \emph{salad} as a main dish would not be satisfactory for the user -- in this situation, we can extend our basic collaborative filtering with a knowledge-based approach that supports the generation of \emph{bundles} taking, for example, into account upper bounds in terms of the number of calories consumed per day \cite{BELADEV2016193}.

\subsection{Quality Education}

\emph{Ensuring inclusive and equitable quality education and lifelong learning opportunities} requires the inclusion of modern communication technologies as well as corresponding personalization concepts which help to tailor learning contents in such a way that  learners can have a personalized learning experience \cite{Klasnja2015}.

An example of applying group recommender systems in e-learning contexts on the macro level is policy decision making regarding the establishment of a new study program at a university. In such a scenario, alternative study programs could be discussed by a group of responsible stakeholders where each stakeholder can provide related proposals him/herself and can give feedback on the other existing proposals/ideas simply by evaluating the interest dimensions \emph{feasibility} (are the personal resources available for teaching the new courses?) and \emph{interest} (will students be interested in enrolling in the new study program?) (see Table \ref{tab:examplestudyprogrammeselection}). We assume an evaluation scale [1..10] $1$ indicating low and $10$ indicating high feasibility/interest.

\begin{table*}[ht]
\centering \caption{Example group decision setting regarding the establishment of a new study program, for example, Artificial Intelligence (AI). Individual stakeholders $s_i$ give feedback on individual proposals in terms of evaluating the interest dimensions (f)easibility and (i)nterest.}
\begin{tabular}{|c|c|c|c|c|c|c|c|c|c|c|c|c|c|c|c|c|c|c|} 
\hline
  stakeholder   &  \multicolumn{2}{|c|}{AI} & \multicolumn{2}{|c|}{AI and Decision Making} & \multicolumn{2}{|c|}{Data Science}  & \multicolumn{2}{|c|}{AI in Software}       \tabularnewline  \cline{2-9}
                  & f    & i           & f  & i                & f     & i           & f    & i  \tabularnewline  \hline
        $s_1$     & 8    & 6           & 6  & 8                & 8     & 4           & 8    & 8  \tabularnewline  \hline
        $s_2$     & 10   & 9           & 2  & 4                & 8     & 2           & 8    & 8  \tabularnewline  \hline
        $s_3$     & 7    & 7           & 8  & 8                & 4     & 2           & 8    & 9  \tabularnewline  \hline
        $s_4$     & 10   & 10          & 4  & 7                & 3     & 3           & 6    & 7  \tabularnewline  \hline \hline
        AVG       & 8.75 & 8           & 5  & 6.75             & 5.75  & 2.75        & 7.5  & 8  \tabularnewline  \hline
\end{tabular} 
\label{tab:examplestudyprogrammeselection} 
\end{table*}

If we assume an equal importance of the interest dimensions feasibility and interest, the \emph{AI} (\emph{Artifical Intelligence}) study program could be recommended to the stakeholders since  it has the highest average (AVG) evaluation. A more detailed discussion on the utility-based evaluation of alternative solutions (items, products) can be found in \cite{Felfernig2006,FelfernigGroupRecommenderSystems2018}.

On the micro-level, there exist a couple of recommendation approaches supporting the recommendation of learning items \cite{Klasnja2015,Ribeiro2011}. On the one hand, content-based filtering can be applied in situations where new learning items are available for learners who are interested in a longterm learning experience regarding a specific topic. This is similar to news recommendation where news gets recommended to users with a corresponding topic-wise reading preference. In the context of university courses, students can estimate their topic-wise expertise by answering corresponding test questions \cite{Stettingeretal2020}. For those topics with a lower knowledge level, content-based recommendation can be used to recommend topic-specific contents ranked on the basis of their complexity level (see Table \ref{tab:examplelearningcontentrecommendation}). 

\begin{table}[ht]
\centering \caption{Example dataset regarding the correctness of student answers to test questions $q_i$ (1=correct, 0=incorrect answer to a question $q_i$).}
\begin{tabular}{|c|c|c|c|c|c|c|c|c|c|c|c|c|c|c|c|c|c|c|} 
\hline
  student      &  \multicolumn{2}{|c|}{$topic_1$} & \multicolumn{2}{|c|}{$topic_2$} & \multicolumn{2}{|c|}{$topic_3$}  \tabularnewline  \cline{2-7}
               &  $q_1$   & $q_2$                 &  $q_3$ & $q_4$                &  $q_5$ & $q_6$                     \tabularnewline  \hline     
    $s_1$      &  1       & 0                     &  1     & 1                    &  1     & 0                         \tabularnewline  \hline     
    $s_2$      &  1       & 1                     &  0     & 0                    &  1     & 0                         \tabularnewline  \hline     
    $s_3$      &  1       & 0                     &  1     & 0                    &  0     & 0                         \tabularnewline  \hline  \hline
correct (\%)   &  1.0     & 0.33                  & 0.66   & 0.33                 & 0.66    & 0                         \tabularnewline  \hline     
\end{tabular} 
\label{tab:examplelearningcontentrecommendation} 
\end{table}

If we assume that Table \ref{tab:examplelearningcontentrecommendation} is a result of a student pre-test questionnaire, the corresponding correctness shares can be used to rank the questions with regard to their complexity. For questions answered incorrectly, corresponding learning contents can be recommended, for example, by a content-based match between question category names and corresponding content categories. For example, student $s_3$ did not answer any question of $topic_3$ correctly. Consequently, contents related to questions $q_5$ and $q_6$ can be recommended (first, learning contents related to $q_5$ since this appears to be a slightly easier topic when following the \emph{correctness} criteria).

\subsection{Gender Equality}
The underlying goal is to \emph{achieve gender equality and to empower all women and girls}. A major aspect in the context of achieving gender equality is the concept of fairness in terms of a gender-independent equal treatment. In recommender systems, fairness aspects play an  important role in terms of assuring this property with regard to stakeholders \cite{Lietal2023}, for example, in music streaming platforms, musicians are interested in having their songs played and users in maximizing their positive song experience.

We expect the availability of different metrics (criteria) that help to analyze the degree to which fairness aspects have to be taken into account as well as pointing out possibilities to counteract unfair treatments \cite{Stray2021WhatAY,Wuetal2023}. Examples thereof are \emph{equal opportunity} requiring the same share of true positives for individual recommender system users or groups, \emph{envy-freeness} indicating to which extent individual users or groups prefer their recommendations over the recommendations given to other users or groups, and \emph{demographic parity} indicating that recommendations should be similar around an attribute such as \emph{gender} \cite{Wuetal2023}. A simple example of how to measure the equal opportunity parity (on a scale [0..1]) of a job recommender is provided in Formula \ref{eq:fairness}.

\begin{equation}\label{eq:fairness}
    fairness = 1 - |accurracy(male) - accurracy(female)|
\end{equation}

There are different ways of assuring fairness \cite{Sonbolietal2022} ranging from (1) the pre-processing of a dataset on the basis of imputation, (2) the provision of fairness-aware algorithms (e.g., on the basis of integrating fairness into machine learning regularization terms) , and (3) the post-processing of generated recommendations (e.g., on the basis of re-ranking recommendations). An example of assuring fairness in a group recommendation scenario (job candidate selection) is depicted in Table \ref{tab:examplegrouprecommendationfairness}.

\begin{table*}[ht]
\centering \caption{Example of stakeholder-specific evaluations of the qualification of different job applicants.}
\begin{tabular}{|c|c|c|c|c|c|c|c|c|c|c|c|c|c|c|c|c|c|c|} 
\hline
  stakeholder & $candidate_1$ & $candidate_2$ & $candidate_3$ & $candidate_4$  \tabularnewline  \hline  
  $s_1$       &    10         &      5        &       6       &        7       \tabularnewline  \hline  
  $s_2$       &    2          &      7        &       8       &        8       \tabularnewline  \hline 
  $s_3$       &    3          &      7        &       7       &        6       \tabularnewline  \hline  
  $s_4$       &    5          &      8        &       5       &        7       \tabularnewline  \hline \hline
  AVG         &    5.0        &      6.75     &       6.5     &        7.0     \tabularnewline  \hline
\end{tabular} 
\label{tab:examplegrouprecommendationfairness} 
\end{table*}

In the scenario shown in Table \ref{tab:examplegrouprecommendationfairness}, stakeholders $s_i$ are in charge of selecting a person for a specific job. In this context, a basic group recommender system is applied to recommend candidates to the group (on the basis of an \emph{avg} aggregation function). In this example, $candidate_4$ has the best overall evaluation which could make him/her the best candidate, however, there is a strong imbalance with regard to the evaluations of $candidate_1$. For this reason, a final decision should not be taken immediately, but discussions need to be triggered regarding the contradicting evaluations of $candidate_1$. Fairness-awareness in this context means to pro-actively figure out potential issues in the decision making process in order to avoid sub-optimal decisions. An important aspect in the context of assuring fairness is also to introduce transparency into decision processes. For example, \cite{TranFairness2019} compare different group recommender user interfaces (differing in terms of decision process transparency) and corresponding stakeholder behaviors in terms of trying to manipulate decision outcomes. A related result is that transparency can help to counteract decision manipulation and thus to reduce the probability of sub-optimal decisions.

\subsection{Clean Water and Sanitation}

Cornerstones for the \emph{availability of clean water and sanitation} are intelligent systems supporting the planning, implementation, and operation of corresponding technical infrastructures \cite{Mahmoud2013,Magalhaes2019}. 

Water management as a whole heavily relies on knowledge about the location-specific quality of water resources which is highly relevant for decision makers, involved in tasks such as land development planning. To identify relevant locations and also to predict the development of water sources over time,  recommender systems can help to predict, for example, the pH level -- for related details on an example application we refer to \cite{Mahmoud2013}. Related techniques for designing relevant sanitation concepts are also in the need of a decision support able to integrate local decision makers \cite{Magalhaes2019}.

In the context of optimizing household water consumption, recommender systems can be applied to sensitize users in terms of adapting, i.e., reducing their water consumption \cite{Arsene2023}. Table \ref{tab:CF4SensitizingUsers} provides a simple example dataset representing different households with corresponding consumption data. Our assumption in this context is the availability of smart-meter technologies allowing the measurement of water consumptions with individual water devices. 

\begin{table}[ht]
\centering \caption{Simplified household water consumption data as a basis for recommending changes in 
 consumption behavior (for shower, bathtub, toilet, and kitchen, the data describes liter p.a.).}
\begin{tabular}{|c|c|c|c|c|c|c|c|c|c|c|c|c|c|c|c|c|c|c|} 
\hline
  household & adults & children & shower    & bathtub & toilet & kitchen  \tabularnewline  \hline  
  $h_1$     & 2      & 2        & 1.000    & 20.000 & 2.000  & 800      \tabularnewline  \hline  
  $h_2$     & 2      & 0        & 300      & 12.000 & 1.200  & 1.000    \tabularnewline  \hline  
  $h_3$     & 2      & 2        & 4.000    & 30.000 & 2.500  & 1.000    \tabularnewline  \hline  
\end{tabular} 
\label{tab:CF4SensitizingUsers} 
\end{table}

In this example (Table \ref{tab:CF4SensitizingUsers}), despite an equivalent number of persons living in the household, household $h_3$ has a significantly higher water consumption compared to household $h_1$. Household $h_1$ can be regarded as a nearest neighbor of household $h_3$. The corresponding differences in consumption can be used as a basis for generating corresponding explanations \cite{Arsene2023}. Depending on the water device specific differences, recommendations can propose actions such as taking shorter showers, using lower-flow shower-heads, and turning off taps during tooth-brushing \cite{Arsene2023}.

\subsection{Affordable and Clean Energy}

The major related goal is the \emph{provision of affordable, reliable, sustainable, and modern energy for all}. Recommender systems can help in the establishment of related energy provision infrastructures such as wind energy systems with layout planning \cite{Sultanaetal2022} and related performance optimizations \cite{Pinciroli2022,Vaghasiyaetal2017}. Achieving the goal of supporting affordable and clean energy also requires the support of public campaigns that indicate in the form of explanations and argumentations which behavior patterns can help to reduce individual energy consumption which is a major goal of assuring affordable and clean energy \cite{Starkeetal2021}. A similar scenario has already been discussed within the scope of the goal of \emph{good health and wellbeing}, i.e., a recommender system can be applied to personalize related messages. Message personalization requires the availability of basic user data such as type of home (e.g., apartment vs. own house), number of family members, and further information regarding personal energy consumption patterns \cite{Eirinaki2022} and also knowledge about persuasive technologies \cite{AdajiAdisa2022} and effective user interfaces \cite{Starkeetal2017}  to achieve sustainable behavior.

On the level of individual households, energy efficiency can be achieved on the basis of household-specific energy breakdowns \cite{Batraetal2017,HIMEUR2021}. In this context, recommendation techniques of collaborative filtering and matrix factorization can help to predict the energy consumption of households who did not perform a breakdown up to now, for example, for reasons of related costs \cite{Batraetal2017}. Household-specific energy consumption can also be triggered on the basis of comparative and community-based explanations \cite{Petkovetal2011} where the energy saving performance of individual households can be compared to each other indicating personal performances compared to other households. \emph{Norm-based comparisons} are an example thereof: \emph{the majority of similar households show a better energy saving compared to your current savings data}. Furthermore, explanations can refer to energy consumption in the past (\emph{self-comparison} feedback) and indicate  improvement or deterioration.

\subsection{Decent Work and Economic Growth}

The underlying goal is to \emph{promote economic growth, full and productive employment, and decent work for all}. Nowadays, recommender systems can be regarded as a core technology helping 
 to further increase the business value of offered products and services \cite{JannachJugovac2019}. Examples of related measurements are \emph{click-through rates} and \emph{sales/revenue}. However, recommender systems supporting  sustainability development goals have a different focus. For example, the impact of recommender systems on increasing the quality of education can be measured directly in terms of increased knowledge levels of different social groups. Furthermore, the impact of recommender systems in the context of clean energy and energy savings can be measured, for example, in terms of reduced household-wise energy consumption. Consequently, for achieving sustainability goals, evaluation metrics  should be more customer-focused and thus also \emph{consequence-based} compared to  metrics in standard business scenarios.

Recommender systems can also help to improve the quality of work and sustainable growth in terms of supporting different kinds of open innovation processes. Achieving sustainability goals is a central agenda of public administrations and finding relevant acceptable solutions for achieving these goals has to be performed in terms of a participatory innovation and design process \cite{BroccoGroh2009,FelfernigRussetal2004,Shadowenetal2020,SmithIversen2018}. In this context, recommender systems can be applied to support idea generation processes, for example, by recommending ideas to community members interested in similar topics  \cite{Haiba2017}.

Recommender systems are an established technology in different people to people (P2P) recommendation scenarios -- examples thereof are recommending new friends in social networks, recommending business partnerships, and recommending jobs \cite{Gutierrezetal2019,Koprinska2022}. Finding the right job is crucial for a further personal development and a productive employment. In these scenarios, recommender systems support a matchmaking functionality by "connecting" job offers with interested employees. Often, such scenarios are based on content-based recommendation where job descriptions are matched with the interest and qualification profiles of potential candidates. An important issue in these scenarios is the aspect of fairness with regard to both, institutions offering a job and corresponding candidates. From the institution point of view, fairness should be guaranteed with respect to other institutions offering similar jobs, i.e., amount and expertise of contacted candidates should be nearly the same. From the candidates point of view, no overloading should take place, i.e., a specific job offer should not be shared with all potential candidates. Finally, a stable or increasing number of new established enterprises can be regarded as a major indicator of economic growth \cite{Luefetal2020} -- in this context, recommender system can be applied to support investors in better identifying the most relevant investments.

\subsection{Industry, Innovation, and Infrastructure}

The underlying goal is to promote \emph{innovation, sustainable industrialization, and resilient infrastructures}. Industrial applications of recommender systems are many-fold and range from the recommendation of movies \cite{GomezUribe2016}, the recommendation of books \cite{SmithLinden2017}, recommendations in the dating business \cite{Tomitaetal2022}, to the recommendation of airline offers \cite{Dadounetal2021}. Beyond acting as a support of core business processes (e.g., selling books), recommender systems can also act in a supportive role which is often the case with sustainability topics. 

Recommender systems can be applied as a knowledge transfer medium for different industrial segments to indicate possibilities in terms of process improvements and the inclusion of sustainable materials into production processes \cite{WiezoreckChristensen2021}. Identifying sustainability properties of products is often not an easy task  -- examples of such properties are environmental impact, animal welfare, and customer benefits \cite{Tomkinsetal2018}. Due to a lack of easily accessible sustainability information, customers do not always behave as intended, i.e., although interested in sustainability, they take sub-optimal decisions due to the lack of related information. \cite{Tomkinsetal2018} introduce a hybrid recommender system where the item-related sustainability classification is based on  probabilistic soft logic.

Fostering innovation can be supported in various forms -- examples thereof are \emph{innovation processes} where recommender systems provide support in the configuration of innovation teams, i.e., who should work together to achieve specific innovation goals \cite{BroccoGroh2009} and the \emph{process of idea generation} \cite{Haiba2017}. An important aspect in software development is to overcome the barriers of taking into account sustainability aspects in software engineering \cite{Roher2013}. Also in this context, recommender systems can be applied to support project stakeholders with recommendations that are determined depending on the underlying application domain. Similar applications exist in software development, where intelligent source code analysis can help to identify software elements to be adapted, for example, to achieve more efficient runtimes and corresponding CPU usage \cite{Muralidharetal2022}.

\subsection{Reduced Inequalities}

Achieving this objective (\emph{reduce inequality within and among countries}) requires actions such as promoting economic inclusion, direct investments, and fostering mobility and migration  to bridge divides.

On the macro-level, recommender systems can help to figure out new potentials overlooked by countries, that can trigger future economic welfare due to strategic future advantages \cite{Liao2018}. In this line of research, recommender systems can also help to establish new study programs of relevance helping to promote relevant know-how for implementing specific industries. As discussed in \cite{Che2020IntelligentED}, recommender systems can be applied in the context of developing export diversification strategies  resulting in recommended industry/product segments which should be expanded or established in specific countries. Having identified such segments, recommender systems can also be applied to identify a corresponding educational focus indicating which study programs should be emphasized or established in a specific country or a specific region \cite{Tavakoli2022}.

Specifically in the context of fostering mobility and migration, the task of country recommendation becomes increasingly relevant. \cite{MajjodietalCountryRecommender2020} motivate the application of country recommender systems since beginning a new life in a different country is for various reasons a high-involvement and often risky decision. The basic underlying idea is to support country recommendation on the basis of collaborative filtering where preferences of existing emigrants are used to infer relevant countries for potential emigrants. Such a scenario can typically not be supported solely on the basis of collaborative filtering (which relies on medium- and long-term preferences) but must include a knowledge-based recommendation component that takes into account short-term circumstances, for example, changing political situations, which do not allow a corresponding recommendation. This is a typical example of hybrid recommendation, where synergy effects of different recommenders  can be combined in a reasonable fashion \cite{Burke2002HybridRecSys}. 

Fairness aspects play a crucial role in different job recommendation scenarios \cite{Lietal2023}. In such scenarios, job candidates should receive recommendations with a very good fit but at the same time companies offering jobs should be treated equally in terms of amount and quality of proposed candidates. A related simplified recommendation scenario is depicted in Table \ref{tab:fairnessjobrecommendation}. Table \ref{tab:fairnessjobrecommendation} shows individual job candidate / job compatibilities determined, for example, on the basis of content-based recommendation which provides a similarity between a job description and the application material provided by the candidate (in our example, on a scale [$1..10$] -- the higher the better).

\begin{table}[ht]
\centering \caption{Simplified example of taking into account fairness aspects in job recommendation scenarios.}
\begin{tabular}{|c|c|c|c|c|c|c|c|c|c|c|c|c|c|c|c|c|c|c|} 
\hline
  candidate & $job_1$ & $job_2$  & $job_3$  & $job_4$ & $job_5$ & $job_6$  \tabularnewline  \hline  
  $c_1$     & 9       & 9        &   8      & 1       & 8       & 1        \tabularnewline  \hline  
  $c_2$     & 9       & 1        &   7      & 9       & 2       & 7        \tabularnewline  \hline  
  $c_3$     & 2       & 1        &   6      & 8       & 7       & 2        \tabularnewline  \hline   
\end{tabular} 
\label{tab:fairnessjobrecommendation} 
\end{table}

In this setting, different fairness aspects can be taken into account. For example, each candidate should have at least one job offering (see Formula \ref{eq:jobspercandidate}).

\begin{equation}\label{eq:jobspercandidate}
    \forall c \in candidates: \#jobs(c)>0
\end{equation}

Furthermore, there should be at least one candidate for each job offering (see Formula \ref{eq:candidatesperjob}). 

\begin{equation}\label{eq:candidatesperjob}
    \forall j \in jobs: \#candidates(j)>0
\end{equation}

Finally, the recommendation quality should be maximized  where \emph{REC} denotes the set of all proposed job/candidate assignments ($rec \in REC$) and  \emph{maxrating} is the maximum (best) possible candidate/job rating. In this context, the optimization goal is to \emph{minimize} the average distance between candidate/job compatibility evaluations and the maximum possible rating (see Formula \ref{eq:recommendationquality}).

\begin{equation}\label{eq:recommendationquality}
    MIN \leftarrow \frac{\Sigma_{rec \in REC}~ maxrating - rating(rec)}{|REC|}
\end{equation}

A recommendation $REC$ for candidate/job assignments on the basis of the example scenario shown in Table \ref{tab:fairnessjobrecommendation} is presented in Table \ref{tab:fairnessjobrecommendationresult}. 

\begin{table}[ht]
\centering \caption{Recommendations of candidate/job assignments where $1$ (in brackets) indicates that the corresponding assignment is part of the recommendation $REC$.}
\begin{tabular}{|c|c|c|c|c|c|c|c|c|c|c|c|c|c|c|c|c|c|c|} 
\hline
  candidate & $job_1$ & $job_2$  & $job_3$  & $job_4$ & $job_5$ & $job_6$  \tabularnewline  \hline  
  $c_1$     & 9(1)       & 9(1)        &   8(1)      & 1(0)       & 8(1)       & 1(0)        \tabularnewline  \hline  
  $c_2$     & 9(1)       & 1(0)        &   7(0)      & 9(1)       & 2(0)       & 7(1)        \tabularnewline  \hline  
  $c_3$     & 2(0)       & 1(0)        &   6(0)      & 8(1)       & 7(0)       & 2(0)        \tabularnewline  \hline   
\end{tabular} 
\label{tab:fairnessjobrecommendationresult} 
\end{table}

In this example, $REC$ consists of $8$ proposed assignments where candidate $c_1$ is recommended for four jobs ($job_1, job_2, job_3, job_5$), $c_2$ is recommended for three jobs ($job_1, job_4, job_6$), and $c_1$ for one job ($job_4$).

Finally, fairness considerations are also relevant in the context of individuals with  disabilities. Related recommendation approaches support content recommendation \cite{Apostolidisetal2022,Quisietal2018}, recommendation for accessibility and mobility \cite{Brodeala2020,Cardosoetal2015,Tsaietal2022}, activity recommendation \cite{Altulyanetal2019}, and the recommendation of points of interest  \cite{Noemietal2022}.

\subsection{Sustainable Cities and Communities}

The related goal is to \emph{make cities and human settlements inclusive, safe, resilient, and sustainable}. City planners, decision makers, and citizens need to be supported in order to achieve the different goals of sustainable cities and communities. For example, sustainable mobility provides modern commuting systems and facilities on the basis of green infrastructures. Furthermore, in order to assure a smart environment, natural resources need to be preserved.

Recommender systems can support sustainable smart cities on the basis of supporting strategic decision making. Depending on the context of a specific city, different actions need to be taken in order to be able to achieve related sustainable development goals \cite{Bokolo2021}. Helping public stakeholders to achieve related sustainability goals can be supported, for example, on the basis of case-based recommender systems which follow the idea of supporting the identification of similar cases (cities) and on the basis of related measures already completed in similar cities to recommend sustainability-fostering activities for the current city \cite{Banerjee2023}.

In such contexts, recommender systems can support also individuals (e.g., citizens and tourists) in the completion of their tasks and the achievement of their goals. For example, sustainable tourism recommender systems are able to propose relevant points of interest (POI) whilst taking into account aspects such as negative environmental impacts,
local communities, and cultural heritage \cite{Banerjee2023,Khanetal2021,Merinov2023}. Related interventions are needed that assure fairness among multiple stakeholders such as tourists, tourism organizations, local citizens, and environmental aspects such as water quality, air quality, and wildlife. Calculating recommendations in such scenarios requires the integration of optimization methods supporting, for example, the optimization of round trips of individual travel groups, resource balancing in the sense that not too many tourists visit specific sightseeing destinations at the same time (triggering issues in terms of disturbances, environmental pollution, and the scaring of animals) \cite{Merinov2023,Sihotangetal2021}. In such contexts, explanations can help to assure recommendation understandability and to sensitize  stakeholders with regard to sustainability aspects \cite{Banerjee2023}.

\subsection{Responsible Consumption and Production}

The underlying goal is to \emph{ensure sustainable consumption and production patterns}. A challenge in this context is to find ways to achieve environment sustainability and at the same time to trigger economic growth and welfare by making these two factors much more independent, i.e., to "achieve more with less".

Sustainable production is related to the goal of achieving industrial symbioses where cooperations between companies are intensified, for example, with the goal to minimize industrial waste streams and share related knowledge \cite{Capelleveenetal2018}. In such contexts, recommender systems can support individual companies by the recommendation of opportunities in waste marketplaces which in the following could lead to intensified cooperations between companies. In such scenarios, recommender systems must be built in a knowledge-based fashion which helps to assure that the needed knowledge about compatibilities of waste products is available. Such basic recommendations can be enhanced by future recommender systems proposing different types of cooperations based on deep knowledge about the underlying waste chains. We regard this scenario as part of the macro level (in the case that public agencies deliver related recommendations for companies) and on the micro-level, if companies themselves are registered in a public marketplace.

Achieving sustainability goals in the fashion industry \cite{Wuetal2022} requires, for example, to lower the number of returned deliveries and to increase a customers willingness to accept higher prices for higher-quality items. Such goals can be achieved, for example, by providing means to create bundles of items \cite{Lietal2020,WiezoreckChristensen2021} which fit together relieving customers from the burden of performing this task on their own \cite{Zielnicki2019}. In this context, persuasive explanations are needed that help to better motivate customers to choose more sustainable options \cite{Knowlesetal2014}. An important aspect is also to assure solution minimality, i.e., to guarantee that product bundles and complex configurations do not entail unnecessary components \cite{VidalSilvaetal2020}.

\subsection{Climate Action}

The major related challenge is to perform actions with the goal to \emph{combat climate change and direct or indirect impacts thereof}. An important aspect in combating climate change is to empower new types of energy production systems, for example, in terms of prosumer networks where private households can act as solar energy producers and consumers at the same time \cite{Guzzi2022TowardsAR}. Before establishing individual cooperations, it is important to figure out and recommend homogeneous prosumer clusters which then maximize the consumption of the cluster-produced energy and -- at the same time -- minimize the consumption of external energy sources. Recommendations in this context can propose specific clusters in a region of consumers \cite{Guzzi2022TowardsAR}. In related energy saving scenarios, persuasive explanations of recommendations play a central role  since households should be encouraged to reduce energy consumption in a sustainable fashion. \cite{Starkeetal2021} show how such explanations can be designed on the basis of the concepts of \emph{framing} \cite{Tversky1985} where those attributes of a decision alternative are highlighted in a recommender user interface which are related to high $kWh$ savings. One simple possibility of "implementing" framing on the user interface level is to sort recommended items on specific attributes making those items more attractive that score high with regard to this attribute. For example, alternative energy saving measures can be sorted with regard to the amount of $kWh$ savings \cite{Starkeetal2021}. These insights regarding the provision of explanations can also be applied in public services provision when informing citizens about potential energy saving measures. Besides the mentioned energy saving scenarios, such persuasive messaging can also be applied in the context of route recommendation scenarios with the goal to encourage users to choose environmental-friendly routes thus contributing to reduce  pollution due to carbon emissions \cite{Bothos2016}.

On the level of individual households, recommender systems can be applied to assist residents in optimizing energy savings. Supporting such optimizations, is a central capability of constraint-based recommender systems \cite{FelfernigBurke2008} which allow the inclusion of optimization criteria to determine relevant recommendation candidates \cite{Murphyetal2015}. If, for example, power suppliers, support time-dependent flexible pricing conditions, the operation of electric equipment should be optimized on the basis of the pricing models. Furthermore, such constraint-based applications can take into account corresponding regional weather forecasts and conditions to also take into account potential consumptions of energy produced by the household itself thus supporting real-time recommendations and corresponding actions in terms of activating and deactivating a specific heating equipment \cite{Dahihandeetal2020}. An important aspect is also that the recommender has knowledge about the current in-building location of residents. Using such knowledge, can help to further decrease power consumption in  buildings by activating/deactivating electronic equipment in an intelligent fashion \cite{Weietal2018}.

\subsection{Life below Water}

The underlying goal is to \emph{enable a sustainable use of oceans, seas, and marine resources}. The application of artificial intelligence techniques in related fields is progressing, however, there is potential for further machine learning and recommender systems applications \cite{Xuetal2022}.

Water quality and pollution assessment and the development of countermeasures becomes an increasingly relevant issue. Due to limited resources in terms of possible data collections and available datasets, machine learning models need to be developed that serve as a basis for pollution prediction but also the determination of recommendations of relevant counter-measures \cite{Xuetal2022}. In the context of illegal fishing, recommender systems can help to propose effective sequential defender strategies that help to counteract illegal fishing \cite{Fangetal2015}.

A relevant problem directly related to water quality and further environmental conditions is the provision of recommendations for aquacultures (e.g., fish farming), for example, in terms of species suitable for the specific conditions and also in terms of nutrients that should be provided in such contexts \cite{Prabaetal2023}. Related recommender applications can also be applied for further tasks, for example, identification and counteracting fish diseases, remote maintenance of offshore infrastructures, and recommending nutrition plans depending a.o. on estimated weight and size of fishes.

\subsection{Life on Land}

The overall underlying goal is \emph{a sustainable use of terrestrial ecosystems}, for example, in terms of \emph{sustainability in forest management, counteracting desertification, and halting of biodiversity loss}. 

It is important to understand and optimally decide on appropriate crops to be cultivated. Crop recommender systems recommend crops on the basis of land quality and mineral requirements whereas pesticide recommender systems propose a collection of pesticides in order to protect specific crops from diseases \cite{AHMED2021106407,PATEL2020105779}. In the line of sustainability requirements, such systems have to take into account impacts of potentially used treatments (e.g., pesticides), i.e., not solely focusing on maximizing productivity but trying to keep soil characteristics are extremely important for maintaining fertility \cite{AHMED2021106407}. In a broader sense, recommender systems can be applied to support different kinds of precision farming \cite{Wakchaure2023,Thilakarathne2022,Ronzhin2022}.

Furthermore, recommender systems can provide suggestions on how to counteract wildlife poaching which is a serious extinction threat to many animal species and related ecosystems \cite{Nguyenetal2016}. Based on such tools, animal protectors are enabled to analyze and predict poaching activities and to recommend countermeasures on the basis of behavioral models learning from poaching data \cite{Nguyenetal2016, Yangetal2014}. In this context, resource balancing plays an important role since personal resources used for observation activities are extremely limited \cite{Yangetal2014}.

\subsection{Peace, Justice, and Strong Institutions}
The underlying goal is to \emph{promote peaceful societies supporting justice for all on the basis of corresponding effective, accountable, and inclusive institutions}. Law enforcement agencies are aware of the fact that the analysis of networks of co-offenders who committed crimes together is highly relevant in crime investigation \cite{Tayebietal2011}. Manually performing such tasks can be quite inefficient which make it an application scenario for recommender systems: suspects are compared with known co-offending networks and the most relevant ones are shown (recommended) to the law enforcement agency representatives.

In the context of trials, recommender systems can support legal practitioners in the identification of advantageous arguments for an ongoing case \cite{Dhananietal2021}. In practice, documents and further material related to the current case are compared with already "closed" cases on the basis of different text-based similarity metrics. The identified most similar documents are then used as a basis for more detailed analysis steps conducted with the goal of identifying relevant arguments better helping to win acquittal for an accused person \cite{Dhananietal2021,Mandaletal2017}. On the negative side, such content-based recommenders are also applied by different social media and news platforms with the danger of creating so-called "echo-chambers" of misinformation \cite{Sallamietal2023} -- this is also related to the general requirement of considering and minimizing harm in recommenders  \cite{Ekstrand2016FirstDN}.

\subsection{Partnerships for the Goals}

The goal is to \emph{identify global partnerships bringing together various institutions such as governments, private sector, and others that help to better achieve the discussed goals}. A specific task is to assure an increasing support for developing countries to assure an equitable progress for all and also strengthen the path towards sustainability. Identifying and establishing such cooperations can also be supported by recommender systems, for example, people-2-people recommender systems can support the identification of business partners and research cooperations \cite{Huetal2021,Koprinska2022}.

\section{Open Research Issues}\label{section:researchissues}

\emph{Evaluation Metrics for Sustainability}. There exists a plethora of evaluation metrics for recommender systems \cite{Zangerle2022} ranging from (1) data-driven approaches to evaluate the prediction/classification quality, (2) experimental settings evaluating prototype systems with alternative variants of user interfaces and algorithmic approaches, and (3) field studies in real-world settings, for example, on the basis of A/B testing. However, existing evaluation metrics do not focus on specific sustainability aspects, for example, achievements in terms of reduced power consumption, increased share of sustainable items in a user's purchase history, and reduced global $CO_2$ footprint -- a specific related aspect is  to take sustainability aspects into account when selecting and/or implementing recommendation algorithms \cite{Lannelongueetal2023,Spilloetal2023}.

\emph{Nudging for Sustainability}. The way decision alternatives are presented to users has an  impact on the final decisions taken by users. In this context, \emph{nudging} \cite{ThalerSuntein2021} can be defined as any aspect of a choice situation that alters the behavior of a user in a predicable way without forbidding any options. Providing a basis for better choice on the basis of decision support is an important goal to be taken into account \cite{Kroeseetal2015}. Related research already indicates the potential of nudges in various recommender systems supporting sustainability goals \cite{Bothosetal2015,KarlsenAndersen2019,Lehneretal2016,MajjodietalNudging2022}. Successful nudges are often based on decision biases, i.e., decision practices (heuristics) used by humans to often lead to suboptimal decision outcomes. An overview of such decision biases and their role in recommender systems is discussed in \cite{Chenetal2013,Mandletal2011,Lexetal2021,TranetalHumanizedRecSys2021}.

\emph{Contextual Explanations}. Given an infrastructure of intelligent data collection, energy consumption information is directly available and can be used for generating corresponding recommendations. For example, in smart homes the activation of a dishwasher and a washing machine could be delayed due to the fact that a parallel car battery recharging would lead to an additional consumption of external energy resources. In travel scenarios, a recommender system can detect alternative (more sustainable) routes not requiring a car rental. In such scenarios, explanations play an important role and must be contextualized and personalized to attain the maximum impact. Explanation generation for achieving sustainability goals can be regarded as a highly relevant research issue \cite{Starkeetal2021}.

\emph{Consequence-based Explanations}. In the context of recommender systems, explanations can be used to support different goals such as trust and persuasiveness (in terms of increasing the probability that a user will purchase an item) \cite{tintarev2012evaluating}. However, with a few exceptions, existing explanation approaches do not take into account the consequences of "accepting" a recommendation. For example, purchasing a rather expensive \emph{BMW} has specific consequences on the economic situation of a household -- having an expensive car could have an impact on the affordability of holidays or the education quality of children. Specifically in the context of achieving sustainability goals, there is a need to analyze alternatives in terms of the corresponding consequences. For example, explanations can provide information regarding the consequences of not investing into new heating equipment (in terms of $CO_2$ footprint issues as well as in terms of additional costs associated with the old (still installed) heating equipment).

\emph{Constraint-based Recommendation for Sustainability}. Constraint-based approaches are applied in various contexts, for example, the optimization of a households energy consumption strategy \cite{Murphyetal2015}. In the line of the idea of simulating the consequences of financial decisions \cite{FanoKurthChoicePoint2003}, constraint-based recommenders could also be combined with corresponding simulation components that help to visualize the impact of different decisions. For example, sticking with the old heating equipment could have an impact on the overall related costs in the long run. Furthermore, consequences exist on different levels, for example, related simulations could also represent "what-if" scenarios, i.e., what happens to the global warming if a majority of people are not thinking about reducing their $CO_2$ footprint.

\section{Conclusions}\label{section:conclusions}

Sustainability development goals (SDGs) as defined by the United Nations are a call for action to planet protection, ending poverty, and ensuring peace and prosperity. In this article, we have provided an overview of  SDGs and related applications of recommender systems. These systems can be regarded as a core technology of different decision support scenarios and thus play a major role in achieving the mentioned SDGs. In order to assure understandability, we have provided corresponding working examples that show how recommender systems can be applied in different application contexts. Furthermore, with the goal to foster further related research, we have provided a list of research issues in the context of developing recommender systems supporting sustainability goals.

\bibliographystyle{plain}
\bibliography{bibliography}





\section*{Funding}
The presented work has been developed within the TU Graz internal project \textsc{AI4Sustainability}.

\end{document}